\pdfoutput=1
\documentclass[12pt,a4paper]{article}
\usepackage[utf8]{inputenc}
\usepackage[margin=1in]{geometry}
\usepackage{amsmath,amssymb}
\usepackage{titlesec}
\usepackage{booktabs}
\usepackage{microtype}
\usepackage{hyperref}

\hypersetup{
    colorlinks=true,
    linkcolor=blue,
    filecolor=magenta,      
    urlcolor=cyan,
    pdftitle={The Industrialization of Noise},
    pdfpagemode=FullScreen,
}

\titleformat{\section}{\large\bfseries}{\thesection}{1em}{}

\title{\textbf{The Industrialization of Noise:}\\ \Large Why Automated Proof-Solving is the Same Old Dead End}
\author{
  \textbf{Richard DeMillo}\\
  \small Georgia Institute of Technology
  \and
  \textbf{Richard Lipton}\\
  \small Georgia Institute of Technology and \\\small Toyota Technological Institute
}
\date{} 

\usepackage{fancyhdr}
\begin{document}

\maketitle
\thispagestyle{fancy}
\begin{quote}
\small
\textbf{$\Lambda$:} \textit{``How do we know that proof checking is always accurate?''}\\[0.5ex]
\textbf{$\mathrm{E}$:} \textit{``How many times do I have to tell you that we know nothing for certain? Your desire for certainty is making you raise very boring problems---and is blinding you to the interesting ones.''}\\
\raggedleft--- \textsc{Imre Lakatos}, \textit{Proofs and Refutations} (1976)
\end{quote}
\vspace{1.5em}

\begin{abstract}
Nearly fifty years ago, DeMillo, Lipton, and Perlis argued that mathematical proofs are not static, isolated structures of formal logic, but the products of a vibrant, evolving social process of human critique, internalization, and shared intuition.\cite{demillo1979social} In October 2026, OpenAI challenged this thesis by releasing a massive repository of 722 automated mathematical manuscripts, claiming full or partial solutions to 372 longstanding open problems. This paper examines the immediate crisis through the lens of the original ``DLP'' framework. We argue that the sudden, automated overproduction of dense, unreadable proofs represents an industrial pollution of the mathematical ecosystem. By examining the three fatal structural pillars of this automated approach---Specification, Regress, and Utility---we demonstrate that decoupling the act of mechanical ``solving'' from the human social process of ``understanding'' produces a sterile repository of dead logical artifacts. 

The resulting industrialization of unexamined manuscripts is self-fulfilling. Flooding the literature guarantees that the finite supply of human attention will lose the race to separate signal from noise. In this sense, posting a result is not the same act as submitting it to the community processes by which mathematics is vetted, digested, and incorporated into shared knowledge.


\noindent\textbf{2020 ACM CCS:} Computing methodologies $\rightarrow$ Artificial intelligence; Theory of computation $\rightarrow$ Automated reasoning.

\end{abstract}

\vspace{1.5em}

\section{The Return of the Formalist Delusion}
On October 6, 2026, the tech industry attempted a corporate take-over of the mathematical commons. In a single, unprompted document dump, OpenAI ``carpet-bombed'' the academic community with 722 automated manuscripts, claiming full or partial solutions to 372 longstanding open mathematical problems. Spanning seventeen branches of mathematics---from algebraic geometry to number theory---this massive drop was celebrated by its architects as an unprecedented leap in machine reasoning.\cite{openai2026mathematics} 

It is nothing of the sort. Instead, it is an act of industrial pollution of the mathematical ecosystem. 

Nearly fifty years ago, we joined our colleague, Alan Perlis, to publish a warning that fractured the calm surface of early computer science.\cite{demillo1979social} In \textit{Social Processes and Proofs of Theorems and Programs}, we argued against the formal program verification movement by stating a fundamental truth about human knowledge: a mathematical proof is not a sterile, static logical object that exists in isolation. It is a result of an evolving social process. A theorem only graduates to the status of ``believed to be true'' after it has been read, criticized, broken down, generalized, and absorbed into the working culture of human minds. The mathematical community operates as a vital social filter, a filter fueled entirely by a strictly limited, highly precious commodity: \textit{human attention}. We call this principle the \textit{DLP Thesis}.

The underlying tension is not new; as University of Edinburgh sociologist Donald MacKenzie documented in his historical analysis of mechanized proof, our 1979 thesis touched off a decades-long debate over whether proof could ever be decoupled from human social processes \cite{mackenzie2001mechanizing}. What the tech industry presents today as a breakthrough in machine reasoning is merely the latest, highly automated iteration of this formalist delusion.

The OpenAI controversy empirically validates our 1979 thesis, executed at a terrifyingly mechanized scale. By utilizing closed, frontier neural networks to generate hundreds of dense, complex manuscripts overnight, the tech industry has reduced the marginal cost of producing logical arguments to exactly zero. 

But human attention remains stubbornly, beautifully finite. 

When the cost of generating text drops to zero while the cost of verifying, assimilating, and determining its value remains high, a catastrophic asymmetry occurs. OpenAI’s release represents a factional capture of science. It demands that the global mathematical community surrender its own research agendas to act as unpaid, subservient content moderators for a private corporation---spending finite human lifetimes sifting synthetic noise for meaningful  signals. 

Some have enlisted the mass media to sound a warning of impending doom for the mathematical enterprise, equating the profligate production of mathematical manuscripts with AI's productivity-enhancing prospects, which currently roil labor markets and policymakers.\footnote{Steven Strogatz, interview on \textit{The Daily}, \textit{The New York Times}, October 13, 2026; ``The End of Math (as we know it)", \textit{Scientific American}, December 2026/January 2027}

Volatile public sentiment reflecting unease about AI has forced even the field's most prominent advocates of formalization to confront the limits of automated problem-solving. In the wake of the drop, Terence Tao suggested that traditional mathematical practice---what he terms ``Math 1.0''---over-indexed on raw problem-solving at the expense of human exposition and community assimilation \cite{tao2026math20}. That characterization mistakes the rarefied economy of elite prize-hunting for the daily reality of mathematical scholarship. Traditional mathematics does not necessarily treat the first brute-force solution as the finish line; it routinely awards its highest honors to conceptual synthesis and illuminating second proofs, from Erd\H{o}s and Selberg's elementary proof of the prime number theorem---delivered half a century after the theorem was already ``proved''---to Erd\H{o}s's entire ethos of proofs from ``The Book.'' 

Tao rightly warns that dumping un-digested machine solutions ``contaminates'' the mathematical landscape by discouraging humans from exploring alternate, insightful pathways \cite{tao2026math20}. However, where reformist voices see a need to recalibrate toward a ``Math 2.0,'' the DLP framework points to a more fundamental reality: an unmotivated, unreadable machine artifact is not merely an undigested proof---it fails to function as a proof at all.

This asymmetry exposes a core fallacy of automated proof: \textit{correctness is not comprehension}. To defend their release, OpenAI noted that 162 of the papers were machine-verified by Lean code. The implication was clear---if a computer checked the syntax, human social verification is obsolete. 

The uncritical acceptance of this implication reveals a profound misunderstanding of mathematical knowledge. Mathematics is not simply a list of symbolic expressions. A Lean-verified proof is a monument built in a desert; it may ensure that a string of symbols is free of logical errors, but it cannot supply a motivated narrative. It cannot tell a human mathematician \textit{why} a particular pathway was chosen out of an infinite landscape of possibilities. It cannot offer the conceptual metaphors that allow a graduate student to apply a technique from geometry to an anomaly in number theory. And it leaves out the false starts, inadequate definitions, counterexamples, and relentless questioning of assumptions that have driven mathematical progress for centuries. As evidence of the eternally tentative nature of mathematical knowledge, one needs to look no further than daily mathematical practice and  the examples we cite below of mathematical innovation spurred by alleged ``proofs" of false statements and true statements that defied formal proof.

Worse yet, a Lean-verified proof fails the tests that G. H. Hardy posed a century ago: 
\begin{quote}
    \textit{``There is, strictly, no such thing as mathematical proof; that we can do nothing but point; that proofs are what Littlewood and I call gas, rhetorical flourishes designed to affect psychology, pictures on the board in the lectures, devices to stimulate the imagination of pupils."}\cite{hardy2004proof} 
\end{quote}

A formal proof---verified or not---does not move a colleague to throw off today's plans for grading homework and commandeer a whiteboard. A proof is a scribble on a coffee-stained napkin, neither correct nor incorrect, but an invitation to a conversation that can go on well into the night. 

A proof is none of the things that a Lean-verified proof purports to be. A proof is, as DLP argued, a message. 
It is not a beautiful abstract object with an independent
existence. No mathematician grasps a proof, sits back, and sighs happily at the knowledge that he can
now be certain of the truth of his theorem. As Yuri Manin observed, ``a good proof makes us wiser."\cite{manin1998goodproofs}

Like Bertrand Russell and Alfred North Whitehead’s disastrous \textit{Principia Mathematica}\cite{russell1910principia}, a proof that is verified by a machine but unreadable by a human is an intellectual dead end. To computationalists like us, a machine-verified proof is a paltry source of information, capable of transmitting a single bit: $1 = \textit{VERIFIED}$ or $0 =\textit{NOT-VERIFIED}.$ It gives us an answer while completely withholding understanding, optimizing humanity out of the equation. 

Mathematics is a body of knowledge that human beings have accumulated, curated, and assimilated over centuries. The formalists' delusion is to face that knowledge and see only dense symbols, unexamined and stripped of humanity.

\section{The Pillar of Specification: Who Checks the Definitions?}
The defense of automated mathematics rests entirely on a sleight of hand. Formalists present a Lean-verified proof as a flawless triumph of pure logic, completely ignoring the messy translation layer that precedes it. Before a frontier neural network can generate a Lean-verified proof, it must first perform an act of radical reductionism by translating a historical, natural-language mathematical conjecture into a rigid, formalized machine specification. 

How that happens has nothing to do with mechanized proof. Euler's formula, $V-E+F =2$ relating the number of vertices, edges, and faces of a solid bounded by flat faces, was proved, checked, and verified using what Euler understood to be the definition of a polyhedron, the simplest of all Archimedean or Platonic geometric objects. He assumed it would be convex, single-piece, and without holes or internal cavities, because the concepts of topology, genus, and convexity did not yet exist in 1750. Euler's formula was proved. Still, it was incorrect. He had no formal way of excluding ``monsters'' like a toroidal picture frame ($V-E+F=0$) or a cube with an internal cavity ($V-E+F=4$). It took another hundred years of human debate, counterexamples, and ``monster-barring'' to produce the rigorous definition of a spherical, convex polyhedron.

The task of checking each definition for monsters and exceptions for Lean proofs does not simply disappear because there is a formal definition that is claimed to capture the informal object being studied. In fact, a supposed advantage of mechanized proofs is their ability to reach across disciplines and make connections hitherto unstudied. But that apparent advantage conceals a trap.  The number of previously unimagined interactions between intuitively similar specifications explodes combinatorially, and the proposition whose truth is at stake must simply wait its turn, unresolved, in mathematical limbo, while more fundamental definitional issues are sorted out. 

Here, the industrialization of noise reveals its first failure mode. A specification complex enough to capture an open problem in pure mathematics is itself a massive, delicate software artifact. It is inherently prone to ambiguity, hidden assumptions, and structural corruption. If the AI model subtly misinterprets the boundary conditions of a conjecture while writing its Lean definitions, it can---and will---produce a ``flawless'' machine proof of a completely wrong, trivialized, or irrelevant version of the problem. MacKenzie's history contains an account of the near-catastrophic consequences of machine-verified proofs based on axioms with undiscovered inconsistencies.\cite{mackenzie2001mechanizing} 

In reality, the dump suffers from the same fatal disease that killed the formal program proving initiatives in the 1980s. The constant refinement by proofs and refutations of underlying assumptions and definitions will---for truly interesting theorems---never result in a clean bill of health. Formal provers will always report \textit{FAILED}, and because the partial proofs are literally unreadable, mathematicians will be left in the dark---a well-documented outcome observed in the relatively primitive verifiers of the 1980s whose set theory axiomatizations were either false in any set theory that included the empty set or inconsistent. Like the software verification systems, there will be no ongoing conversation about how to fix the bugs.

We already see this breakdown in OpenAI’s October dump.  Human mathematicians are finding that they cannot simply audit the proofs; they must first spend days interrogating the definitions to see if the AI even understood the question it claimed to solve.  When a private corporation floods the ecosystem with hundreds of papers, it forces the academic community to waste its finite cognitive resources verifying the machine’s vocabulary.  A mathematically perfect proof of a corrupted specification is not progress; it is noise masquerading as certainty.

\section{The Pillar of Regress: Who Checks the Checker?}
The second Formalist pillar is the belief that formal verification halts the infinite regress of human skepticism.  \textit{“The code compiles, therefore the argument is settled.”} This claim ignores the foundational reality of computer science: a formal verifier is not an ethereal oracle; it is a piece of software written by fallible humans, running on an engineering stack riddled with hidden dependencies.

To accept OpenAI’s 162 machine-verified papers, a mathematician must accept a dizzying, multi-layered game of digital telephone.  You must trust the proprietary, black-box LLM reasoning chain that generated the argument.  You must trust the translation layer that converted those neural tokens into Lean code.  You must trust the specific version of the Lean compiler used to run the verification kernel.  And finally, you must trust that the human engineers who wrote the Lean kernel did not introduce a single catastrophic bug into its thousands of lines of source code. 

By automating proof generation, the tech industry has not eliminated the need for trust; it has merely obscured it behind an uninspectable wall of software infrastructure. When OpenAI was forced to quietly withdraw and revise several of its mathematical manuscripts shortly after their release due to ``errors,'' they exposed the  reality we predicted fifty years ago: the automated pipeline itself is a fragile, buggy construct. The infinite regress remains unbroken. If human mathematicians cannot trace, critique, and verify every link in the chain---including the software tools themselves---then the ``proof'' is nothing more than a dogmatic assertion wrapped in a compiler script.

Computer science recognized this infinite regress decades ago.  In his classic 1984 reflection on software verification, Ken Thompson demonstrated that you cannot trust code simply because you can inspect it, because the tools used to inspect, compile, and execute that code rest on prior layers of software that must themselves be trusted blindly \cite{thompson1984reflections}. When applied to automated mathematics, Thompson's paradox is inescapable: to trust a Lean-verified proof of a hard conjecture, one must trust the Lean kernel, the C++ compiler that built the kernel, the operating system beneath it, and the neural model's token-translation layer. The formalist claim that code halts skepticism merely hides the human chain of trust beneath layers of compiled abstractions. Thompson's warning was prophetic.

The Lean home page falsely promises certainty by claiming ``Lean's minimal trusted kernel guarantees absolute correctness in mathematical proof, software and hardware verification."\footnote{\url{https://lean-lang.org/}}  However, fragile reality is starkly different, as Terence Tao laid bare in what he characterized as the ``summer of Lean soundness bugs.'' \cite{tao2026lean}  Rather than discovering subtle mathematics, automated tools and AI agents turned loose on the proof assistant exposed systemic vulnerabilities in the software verifier itself. Using Trusted Computing Base (TCB)-Expansion attacks, adversarial prompts managed to smuggle unverified code and false axioms past Lean's auditing commands \cite{meza2026tcb}. 

The consequences were immediate and glaring: invalid AI-generated proofs—including a widely publicized disproof of the Collatz conjecture—were granted clean, verified status by the Lean kernel. \cite{paulson2026kernel}  Crucially, independent third-party checkers like Nanoda suffered a coincidental collapse, failing to detect the errors due to shared architectural blind spots \cite{paulson2026kernel}. This crisis revealed a vital truth: synthetic proof generation doesn't just clutter archives with unread text; at scale, it stress-tests, corrupts, and breaks the very software checkers meant to guarantee mathematical truth.

At scale, the Summer of Soundness Bugs exposed the same deep flaws in the Lean agenda that infected formal program verification: a very small, heavily audited piece of code (the trusted kernel) checks all proofs. Once the kernel signs off on a proof, it achieves absolute correctness status and becomes ground truth for all subsequent proofs. But the sheer volume of generated proofs guarantees that most flaws will go undetected.  

The vanishing signal-to-noise ratio carries a structural danger that strikes at the very heart of the formalist program: the silent corruption of the mathematical foundation itself. The foundationalists promised an unshakable fortress of truth, free from human error. Yet in any formal logical system, an undetected flaw is not a minor nuisance. It is a catastrophic breach. By the logical principle of \textit{ex falso quodlibet}, a single false proposition or unsound axiom smuggled into a core library allows any arbitrary statement, true or false, to be formally verified. As automated systems flood the mathematical ecosystem with millions of synthetic artifacts that no human mind will ever read or audit, the opportunities for undetected defects to enter and propagate through shared foundational libraries multiply.

This is how the industrialization of proof introduces a systemic vulnerability we call \textit{Single Point of Collapse}. As machine-verified lemmas become dependencies for ever-larger bodies of formal mathematics, a small, undetected defect can have an arbitrarily large downstream impact. A plausible counter-argument is that the stakes are so high that future, unspecified architectural innovations can be counted on to eliminate such flaws. Maybe so, but, as Thompson pointed out\cite{thompson1984reflections}, that assumes the machines themselves are benevolent. Relying on automated verifiers to police machine-generated proofs merely shifts the trust bottleneck to complex software pipelines that autonomous vulnerability-discovery systems have already proven capable of exploiting\cite{kim2025atlantis}. 

The resulting concentration of mathematical trust creates a fragile foundation beneath an expanding universe of apparently verified results. Far from securing the discipline, the unvetted industrialization of noise threatens to transform machine-checked mathematics into a house of cards, where vast towers of formal proof rest unknowingly on an inconsistent foundation.

\section{The Pillar of Utility: Is a Blind Proof Useful?}
Let us grant the Formalists their most radical, impossible assumption. Let us assume that the specification is perfect, the Lean compiler is flawless, and the proof-object is entirely free of logical errors. We are still left with this question: \textit{So what?} 

Of what earthly use is the knowledge that a proof \textit{exists} if that proof is unreadable? In 1979, we argued that the utility of a mathematical proof does not come from a final, binary stamp of correctness. It comes from its role as a social vehicle. A proof is valuable because it is an instrument of pedagogy and discovery; it must be internalized by other minds so that its novel techniques, its elegant shortcuts, and its conceptual breakthroughs can be stolen, adapted, and used to solve entirely different problems across science. As a practical matter, no one systematically edits mathematical knowledge by detaching \textit{if} clauses, so truly interesting mathematical knowledge conditioned on a conjecture remains valid regardless of the truth of the conjecture. The history of mathematics is replete with recorded proof assimilation in textbooks, biographies, and even in \textit{prior work} sections of manuscripts. And sometimes a proof is valuable because it is wrong, and the process of reading and refuting it leads to new mathematics.

An AI model optimizing millions of tokens within a closed logic loop does not produce a conceptual toolkit. It produces a dense, unreadable maze of brute-force axiomatic transitions. It provides an answer while  destroying the path that led there. If a graduate student cannot read an OpenAI manuscript and extract a new mental model from it, then the utility of that manuscript is effectively zero. It does not advance human capability to think; it freezes mathematics in its tracks, turning a living, conversational science into a dead repository of private corporate data. Knowing that a problem has been ``solved'' by a machine gives humanity no more insight into the universe than knowing that a supercomputer can calculate $\pi$ to the trillionth digit. 

\section{GitHub is Not an Institution: The AMS Test}

The fundamental error of the automated formalist push is a confusion of entitlement. Proponents of automated theorem provers act as if the mere existence of a Lean-verified code repository creates a moral obligation for the mathematical community to inspect, verify, and assimilate it. It does not. A GitHub push is not a scientific submission, and GitHub is not an institution. 

AI researchers often object to a social process that seems biased against automated output. This is a strange grievance coming from a group that feels empowered to obligate the global mathematical community to take hundreds of claimed results seriously without having subjected a single one to meaningful human examination. The gateway to mathematical knowledge is, and has always been, open to all contributors. But no amount of automated proof production can manufacture the social acceptance on which mathematical knowledge depends. The burden of demonstrating mathematical significance belongs entirely to the producer of a result, not to the community receiving it.

The mathematical profession already possesses established institutional mechanisms to handle this  dynamic. The American Mathematical Society (AMS), for example, has a longstanding tradition of accepting reasonable contributed-paper abstracts for presentation at its national and regional meetings. As AMS associate secretary Steven H. Weintraub reflected in the May 2025 \textit{Notices of the AMS}: ``The AMS policy is that we accept all reasonable contributed paper abstracts'' \cite{weintraub2025ams}. The associate secretary screens submissions primarily to eliminate mathematically unintelligible papers, elementary ``proofs'' of Fermat's Last Theorem, or naive trisections of the angle. While recent guidelines for meetings such as the Joint Mathematics Meetings retain screening and acceptance notifications, the underlying core policy remains intact: acceptance for presentation is not a certificate of correctness, but an invitation to professional discourse.

This tradition provides a practical, real-world test of the DLP thesis using the mathematical profession's own procedures. The challenge to developers of automated theorem provers is simple: If your AI has solved an important mathematical problem, submit an abstract to an AMS meeting and present the result to an audience of mathematicians.

Choose a theorem. Submit an abstract. Send a representative who can explain the result, defend its formal definitions, answer spontaneous questions from the floor, and persuade working mathematicians that the argument deserves their finite attention. There is no need to flood the community with 722 manuscripts. One genuinely interesting result will do. If the argument survives that human encounter, mathematics will have gained something. If it fails, mathematics will have lost nothing.

This institutional design separates access to discourse from certification of truth. But it also imposes a severe constraint that corporate AI labs have so far refused to face: reputational risk. Presenting manuscripts that turn out to be mathematically incoherent, defectively specified, or logically broken at AMS sessions eventually inflicts lasting reputational damage on the author. That reputational penalty applies to corporate authors just as heavily as to individual academics. 

It is telling that the 722 manuscripts released in OpenAI's October 2026 dump bypassed the established forums of mathematical discourse. The October dump reportedly contains multiple mathematical defects. Had those results been presented and defended through standard professional channels, the resulting failures could have damaged the reputational standing of both the corporation and its automated verification pipeline. 

The decision to execute an unprompted document dump on GitHub rather than engaging established mathematical forums was not an act of scientific confidence; it was an evasion of institutional accountability. That realization seems to have occurred to corporate sponsors. Selected manuscripts are now reportedly being prescreened by expert ``referees," but the use of hand-picked reviewers who know both the provenance of a submission and the expected outcome of the review is a flagrant violation of ethical norms for peer-reviewed publications. 

The mathematical community has filtered claims through face-to-face scrutiny for generations. It has no obligation to abandon that practice merely because the marginal cost of generating purported proofs has approached zero.

\section{Conclusion: The Resilience of the Social Process}
The Formalist agenda has been recognized as a practical and conceptual dead end for over half a century. The sudden arrival of generative AI has not magically reanimated that agenda; it has merely motorized it, enabling the large-scale industrialization of noise. 

Mechanized proof demands neither panic nor passive surrender. Mathematicians are not helpless bystanders to a corporate token dump. The mathematical enterprise has never been regulated by the sheer volume of papers uploaded to archives, nor is it bound to award credit to raw, unmotivated logical execution. Because human attention is strictly finite, the economics of the mathematical ecosystem remain firmly in the hands of the community. 

We do not seek to outlaw Lean-verified proofs any more than, say, long, boring calculations or tedious, machine-assisted case analysis. On the other hand, if an automated system occasionally produces a fragment of genuine insight---a novel lemma or a missing link buried in thousands of pages of synthetic output---that fragment does not automatically bypass the historic filters. To become mathematics, it must still be extracted, subjected to peer critique, stripped of its mechanical bloat, and integrated into human understanding. 

At its core,  mathematics is not merely an enterprise for accumulating verified proofs. It is, above all, the search for the right definitions, the right concepts, and the right ideas. 

Consider the Riemann Hypothesis: systematically improving the verified numerical bounds of non-trivial zeroes along the critical line—say, pushing the verified proportion from forty percent to sixty percent—represents an impressive feat of computational engineering. Yet that result pales in comparison to Bernhard Riemann’s original, audacious conjecture that \textit{all} non-trivial zeroes lie on that line. The monumental leap was not a brute-force verification of instances, but the invention of a conceptual framework that redefined number theory for centuries.

Whatever one thinks of the DLP argument, mathematicians are all bound by the Socratic obligation : ``Agree with me if I seem to speak the truth." Socrates asked for assent conditioned on the merits of the argument, not on the authority of the speaker.  Agreement is the consequence of examination, not its starting point.

Machine-generated proofs, Lean-verifications, DLP examinations in the public square, and subsequent mathematical applications all offer different grounds for rational inference.  None should acquire epistemic authority on the basis of provenance alone.

This brings us full circle to the DLP Thesis and Imre Lakatos’s insight in \textit{Proofs and Refutations}: proofs are not final trophies or static binary certificates \cite{lakatos1976proofs}. They are intellectual vehicles whose true purpose is to sharpen our concepts and expand human intuition. 

Corporate AI labs can automate the generation of syntax, but they cannot automate the social assignment of meaning. The industrialization of noise will produce vast mountains of unread artifacts, but the living field of mathematics will continue as it always has—driven by human curiosity, governed by finite attention, and measured by our shared capacity to find the right ideas.
\bibliographystyle{plain}
\bibliography{industrialization_of_noise}

@article{demillo1979social,
  author    = {DeMillo, Richard A. and Lipton, Richard J. and Perlis, Alan J.},
  title     = {Social Processes and Proofs of Theorems and Programs},
  journal   = {Communications of the ACM},
  volume    = {22},
  number    = {5},
  pages     = {271--280},
  year      = {1979},
  publisher = {ACM New York, NY, USA},
  doi       = {10.1145/359104.359106}
}

@book{lakatos1976proofs,
  author    = {Lakatos, Imre},
  title     = {Proofs and Refutations: The Logic of Mathematical Discovery},
  year      = {1976},
  publisher = {Cambridge University Press},
  address   = {Cambridge}
}

@book{russell1910principia,
  author    = {Russell, Bertrand and Whitehead, Alfred North},
  title     = {Principia Mathematica},
  volume    = {1},
  year      = {1910},
  publisher = {Cambridge University Press},
  address   = {Cambridge}
}

@misc{openai2026mathematics,
  author    = {{OpenAI}},
  title     = {Sharing AI Progress in Mathematics},
  howpublished = {OpenAI Blog},
  year      = {2026},
  month     = {October},
  note      = {Released October 6, 2026},
  url       = {https://openai.com/index/sharing-ai-progress-in-mathematics/}
}

@incollection{hardy2004proof,
  author    = {Hardy, G. H.},
  title     = {Mathematical Proof},
  booktitle = {Musings of the Masters: An Anthology of Mathematical Reflections},
  editor    = {Ayoub, Raymond G.},
  series    = {Spectrum},
  volume    = {42},
  publisher = {Mathematical Association of America},
  year      = {2004},
  isbn      = {978-0-88385-549-2}
}

@misc{tao2026math20,
  author = {Tao, Terence},
  title = {Math 1.0 vs Math 2.0 and the Contamination of Problem Spaces},
  journal = {Mathstodon},
  month = oct,
  year = {2026},
  url = {https://mathstodon.xyz/@tao}
}

@book{mackenzie2001mechanizing,
  author    = {MacKenzie, Donald},
  title     = {Mechanizing Proof: Computing, Risk, and Trust},
  year      = {2001},
  publisher = {MIT Press},
  address   = {Cambridge, MA}
}

@article{thompson1984reflections,
  author  = {Thompson, Ken},
  title   = {Reflections on Trusting Trust},
  journal = {Communications of the ACM},
  volume  = {27},
  number  = {8},
  pages   = {761--763},
  year    = {1984}
}

@article{manin1998goodproofs,
  author       = {Manin, Yuri I.},
  title        = {``{G}ood proofs are proofs that make us wiser''},
  subtitle     = {Interview with {Y}uri {I}. {M}anin},
  editor       = {Aigner, Martin and Schmidt, Vasco A.},
  journal = {Mitteilungen der Deutschen Mathematiker-Vereinigung},
  volume       = {6},
  number       = {2},
  year         = {1998},
  pages        = {40--44},
  note         = {Interview conducted by Martin Aigner and Vasco A. Schmidt}
}

@article{weintraub2025ams,
  author  = {Weintraub, Steven H.},
  title   = {Reflections on Sixteen Years as an {AMS} Associate Secretary},
  journal = {Notices of the American Mathematical Society},
  volume  = {72},
  number  = {5},
  pages   = {580--583},
  year    = {2025},
  month   = 05
}

@misc{tao2026lean,
  author = {Tao, Terence},
  title  = {What mathematicians should know about the {Lean} {T}heorem {P}rover (questions of reliability and {AI})},
  journal = {What's new},
  month  = 10,
  year   = {2026},
  url    = {https://terrytao.wordpress.com/2026/10/09/what-mathematicians-should-know-about-the-lean-theorem-proverquestions-of-reliability-and-ai/}
}

@misc{paulson2026kernel,
  author = {Paulson, Lawrence C.},
  title  = {Why is it all in the kernel?},
  journal = {Machine Logic},
  month  = 06,
  year   = {2026},
  url    = {https://lawrencecpaulson.github.io/2026/07/30/Collatz.html}
}

@techreport{meza2026tcb,
  author      = {Meza, Juliet},
  title       = {{TCB}-Expansion Attacks on {Lean 4} and the {LLM} Proof Reviewers That ({M}ostly) Miss Them},
  institution = {Apart Research},
  year        = {2026},
  month       = 05,
  url         = {https://apartresearch.com}
}

@misc{kim2025atlantis,
  author       = {Kim, Taesoo and Han, Hyungseok and Park, Soyeon and others},
  title        = {{ATLANTIS}: {AI}-driven Threat Localization, Analysis, and Triage Intelligence System},
  year         = {2025},
  eprint       = {2509.14589},
  archivePrefix = {arXiv},
  primaryClass = {cs.CR},
  url          = {https://arxiv.org/abs/2509.14589}
}
\end{document}